\documentclass[12pt,fleqn]{article}
\usepackage[dvips]{color}
\usepackage[dvips]{graphicx}

\begin{document}

\title{Comment on: Direct space-time observation of pulse tunneling in an electromagnetic band gap,
by S.Doiron, A.Hache, H.Winful~\cite{Winful}}

\author{G. $Nimtz^{1,2}$ and A.A. $Stahlhofen^2$\\
$^1$II. Physikalisches Institut, Universit\"at zu K\"oln and\\
$^2$Institut f\"ur Integrierte Naturwissenschaft, Universit\"at
Koblenz}

\date{}
\maketitle

The title of this article is misleading. The authors have
investigated a resonator but not a tunneling barrier see also
Refs.\cite{Winful2} The measured superluminal group velocity and
discussed is that studied on a Lorentz-Lorenz oscillator by
Sommerfeld and Brillouin a hundred years ago~\cite{Brillouin}. It
is similar to the faster than light experiment by Wang et al.
based also on anomalous dispersion with a complex refractive index
of a resonator ~\cite{Wang}.

Tunneling, however, is understood and performed by electromagnetic
evanescent modes or by tunneling solutions of the Schr\"odinger
equation, which have purely imaginary wave numbers. The latter
includes a purely imaginary refractive index. Signals with purely
evanescent frequency components can travel at a superluminal
velocity ~\cite{NimtzH,Nimtz1}. Inside the barrier tunneling
proceeds even instantaneously, i.e.by a process described by
virtual photons~\cite{Stahlhofen}.

Actually, in the paper there are some errors: Fig.3 shows the
vacuum light velocity and in section (II,D) the dwell time is not
directly measured  but it is derived from an approximately
integrated stored energy and from the measured input power. In
addition it is claimed to have measured a resonator decay time,
but detectors measure the traversal time of a black box
independent of the content of the box. The authors are asking
\emph{whether an identifiable pulse peak actually propagates
through the barrier?} According to their Fig.1 not only the peak
but also the pulse half width (for instance representing a digital
signal) propagated faster than light and were correctly detected.
Remember a signal does not depend on its magnitude as long as it
is above the noise level.


\end{document}